\documentclass[twocolumn]{jpsj3}
\usepackage{graphicx,amsmath,amssymb,bm}
\usepackage{amsfonts}

\allowdisplaybreaks 

\usepackage{color}
\definecolor{Green}{rgb}{0,0.7,0}

\newcommand{\ve}{ \bm{e}}
\newcommand{\vv}{ \bm{v}}
\newcommand{\vq}{ \bm{q}}
\newcommand{\vk}{ \bm{k}}

\newcommand{\e}{ {\rm e}}

\newcommand{\bk}{ \bm{k}}
\newcommand{\bq}{ \bm{q}}
\newcommand{\bkD}{ \bm{k}_{\rm 0}}
\newcommand{\kDx}{ k_{\rm 0x}}
\newcommand{\kDy}{ k_{\rm 0y}}
\newcommand{\kDz}{ k_{\rm 0z}}
\newcommand{\tq}{ \tilde{q}}
\newcommand{\td}{ \vec{\delta}}

\begin{document}
\title{
Role of Velocity Field and Principal Axis of Tilted Dirac Cones 
 in Effective Hamiltonian  of Non-Coplanar Nodal Loop
}

\author{
 Yoshikazu Suzumura$^{1}$\thanks{E-mail: suzumura@s.phys.nagoya-u.ac.jp}
Takao Tsumuraya$^{2}$, 
Reizo Kato$^{3}$, 
 Hiroyasu Matsuura$^{4}$, and 
 Masao Ogata$^{4}$
}
\inst{
$^1$
Department of Physics, Nagoya University,  Nagoya 464-8602, Japan \\
$^2$
Priority Organization for Innovation and Excellence, 
Kumamoto University, 
 Kumamoto~860-8555, Japan \\
$^3$
 Condensed Molecular Materials Laboratory, RIKEN,  Wako, Saitama 351-0198, Japan \\
 $^4$
Department of Physics, University of Tokyo, Bunkyo, Tokyo 113-0033, Japan \\
}
\recdate{May   \;\;\;, 2019}
\recdate{ \;\;\;\;\;\;\;\;}

\abst{
 A nodal line  
   in a single-component molecular conductor [Pd(dddt)$_2$] 
  with a half-filled band has been examined 
    to elucidate the properties of a Dirac cone on the non-coplanar loop. 
 The velocity of the tilted cone  is evaluated  
   at respective Dirac points on the nodal loop, which is obtained by
        our first-principles band structure calculations~[J. Phys. Soc. Jpn. {\bf 87}, 113701 (2018)].
        In the previous study, we proposed a new method of deriving 
 an effective Hamiltonian with a 2 $\times$ 2 matrix using 
two kinds of velocity of the Dirac cone on the nodal line, by which 
  the momentum dependence of the Dirac points is fully reproduced 
 only at symmetric points. 
In this work, we show that our improved method 
 well  reproduces  reasonable behavior of all the Dirac cones and 
 a very small energy dispersion of 6~meV among the Dirac points on the nodal 
 line,  which originates from the three-dimensionality of the electronic state.
 The variation of velocities along the nodal line is  shown 
by using the principal axes of the gap 
       between the conduction and valence bands. 
Furthermore, such an effective Hamiltonian is applied to 
calculate the density of states close to the chemical potential 
and the orbital magnetic susceptibility.
}

\maketitle

\section{Introduction}

The class of three-dimensional~(3D) topological semimetals called nodal line semimetals is a recent topic in condensed matter physics.
\cite{Murakami2007,Burkov2011,Fang2016,Hirayama2018,Bernevig2018,Fang2015} 
Although a number of band calculations have predicted the existence of nodal line semimetals near the Fermi level\cite{Kim_Rappe2015, Mullen2015, Yu_Cu3PdN2015, Yamakage_CaAgX_Dirac, Huang2016R, Xie2015, Quan2017, Hirayama2017, Matsuura2018}, only a few candidate materials have been  experimentally confirmed by angle-resolved photoemission and magnetoresistance.\cite{Okamoto2016, Schoop2016, Wang_CaAgAs2017, Takane_CaAgAs2018}
There are several  protection mechanisms  of   nodal line  against vanishing,  
 such as   
 a combination of inversion and time-reversal symmetry, mirror reflection symmetry, and nonsymmorphic symmetry.\cite{Yang2018}
The nodal line takes the form of an extended line running across the Brillouin zone (BZ), a closed loop inside the BZ 
 or even  a chain  of tangled loops. 
 Such forms originate from accidental degeneracies in  energy bands 
 with an inversion symmetry~\cite{Herring_PR.52.365}.
The existence of an odd or even number of nodal loops inside the BZ corresponds
  to the condition of a   negative or positive sign
    of a product of parity eigenvalues of filled bands 
   at the time-reversal-invariant momentum (TRIM), respectively.  
This condition is also valid  
   for weak spin--orbit coupling (SOC) materials with light elements 
 such as molecular conductors. 
    The classification of band nodes has been recognized 
 as underpinning topological materials since the discovery of the 
$\mathbb{Z}_2$ topological insulator.
\cite{Fu_Kane2007, Fu_Kane_Mele2007, Song_Fang_Z4}

A notable  molecular conductor that shows a single nodal-loop semimetal was discovered   
  by  first-principles calculation  and transport measurement under pressure.  
 A single-component molecular conductor [Pd(dddt)$_2$] (dddt = 5,6-dihydro-1,4-dithiin- 2,3-dithiolate) exhibits nearly massless Dirac electrons under high pressure, 
 as shown by its almost temperature-independent electronic resistivity 
  and by theoretical structural optimization 
    using first-principles calculations 
    based on density functional theory~(DFT).~\cite{Kato_JACS}
Furthermore, the nodal line with a loop of Dirac points has been 
analyzed using an extended H\"uckel calculation for the DFT-optimized structure.~\cite{Kato2017_JPSJ} 
The formation of Dirac points originates from the multiorbital nature, where the parity is different 
  between   
    the  highest occupied molecular orbital (HOMO) and 
  the lowest unoccupied molecular orbital (LUMO).
   
The characteristic  property of the nodal line semimetal has been  examined to comprehend such a nodal line.   
 We have calculated the anisotropic electric conductivity at absolute zero 
  and finite temperatures~\cite{Suzumura_Conduc_2017,Suzumura2018_JPSJ_T}
   and proposed the reduced Hamiltonian 
    with two components.~\cite{Kato2017_JPSJ,Liu2018}
Furthermore, the extensive studies have been  performed on the topological behavior of the Berry phase~\cite{Suzumura_Yamakage_JPSJ} and 
   on a method of obtaining  an effective Hamiltonian directly from the nodal line.~\cite{Tsumuraya2018_JPSJ}  
 To elucidate the condition of the Dirac electrons,~\cite{Fu_Kane2007,Kim_Rappe2015} 
    the present Dirac nodal line semimetal in a 3D system 
      is compared with the previous case of massless Dirac electrons 
      in a two-dimensional molecular conductor.
~\cite{Katayama2006_JPSJ75,Kajita2014,Piechon2013_JPSJ,Kato2017_JPSJ} 
 Note that   [Pd(dddt)$_2$] 
   may be regarded as a Dirac electron system 
    with  a  gapless nodal line,~\cite{Kato_JACS,Tsumuraya2018_JPSJ}
   although it 
      becomes  a strong topological insulator~\cite{Tsumuraya_APS}
        in the presence of  SOC.~\cite{Fu_Kane_Mele2007}

 In the previous work, 
  a reduced model was introduced  to analyze   Dirac cones 
  in  [Pd(dddt)$_2$].~\cite{Tsumuraya2018_JPSJ} 
 In fact,    an effective Hamiltonian with a 2 $\times$ 2 matrix 
    was derived  by employing a new method where 
 two kinds of velocities of the cone are successfully 
     calculated from  the momentum dependence   of the Dirac points on 
       the nodal line.
However,  the  description of the 
 the matrix element  is insufficient  to reproduce  the quantitative 
 behavior of  all the Dirac cones on  the  nodal line. 
 The directions of both the velocity and  the principal axes of the cone  
    are nontrivial, and   the  cone is  tilted  when  
   the energy of the  Dirac point depends on the line. 
 Furthermore, it is significant to determine the principal axes 
 of  the cone 
   to calculate the correct response to the external field, 
 as seen from  the deviation of the current from the electric field for the 
 anisotropic conductivity.~\cite{Suzumura2014}
 
In the present paper, 
by improving the previous  method,~\cite{Tsumuraya2018_JPSJ}
  we  demonstrate   the effective model that reproduces 
    all the  Dirac points obtained in  the DFT calculation.  
In Sect.~2, 
 the velocities of the  Dirac cone are calculated 
    from  the gradient of matrix elements,
      while the tilting velocity is obtained 
       from  the  energy variation  of the Dirac point.
In Sect.~3, the variation of the Dirac cone along the nodal line 
 is examined by calculating the velocity fields and principal axes 
 of the cone, which is  obtained from the gap 
between the conduction  and valence bands.
The effect of  tilting  the cone is shown by calculating 
 a tilting parameter.
In Sect. 4,  using the present effective Hamiltonian,  
   the density of states (DOS)  and orbital magnetic susceptibility 
  are calculated  
       to understand the characteristics of the nodal line semimetal. 
A summary is given in Sect.~5.  

\section{Nodal Line and Two-Band Model}
\begin{figure}
  \centering
\includegraphics[width=7cm]{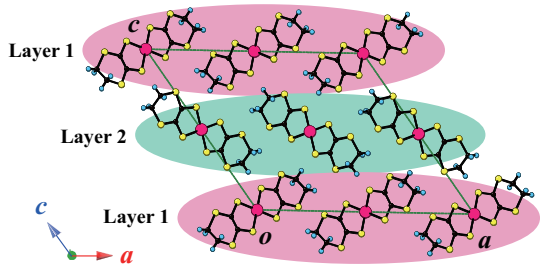}
    \caption{(Color online)
Crystal structure of 
 [Pd(dddt)$_2$] on the plane
  with the $\bm{a}$ and $\bm{c}$ axes.~\cite{Kato_JACS}  
 The most conducting axis is given by $\bm{b}$ 
    being perpendicular to the $\bm{a}$-$\bm{c}$ plane.
 There are four Pd(dddt)$_2$ molecules in the unit cell (the solid line), 
 which consists of two layers shown by Layer 1 and Layer 2. 
 Each molecule on an inversion center at Pd atom 
 has  HOMO and LUMO with the different parity. 
  }
\label{213fig1}
\end{figure}

\subsection{Effective Hamiltonian}
Figure \ref{213fig1} shows the crystal structure of  
 the single-component molecular  conductor [Pd(dddt)$_2$],
 where there are two layers, 1 and 2, that are  crystallographically 
 independent. 
The Dirac point is determined by the HOMO band of  Layer 1 and 
 the LUMO band of  Layer 2.   
 In the molecule, there is an  inversion center at 
   the Pd atom, 
  where the HOMO and LUMO have different parities
   of ungerade and gerade symmetries. 
 Since there are four molecules in the unit cell, 
 there are eight energy bands,  $ E_1 > E_2 > E_3 > \cdots >E_8$, 
    where the upper (lower) four bands  are mainly determined by the  LUMO (HOMO). 
Under a high pressure of 8 GPa, 
  the electronic state  
    shows the Dirac point due to  the reverse given by  
  $E_4(\bk)$ for the HOMO and $E_5(\bk)$ for the LUMO close to the $\Gamma$ point.
The tight-binding model shows 
  that the Dirac points $\bk_0$ with $E_4(\bk_0)=E_5(\bk_0)$
 form a loop, i.e., a nodal line between the conduction and valence  
      bands.~\cite{Kato2017_JPSJ} 
Such a line has been  verified by   first-principles DFT calculation.~\cite{Tsumuraya2018_JPSJ}

\begin{figure}
  \centering
\includegraphics[width=6cm]{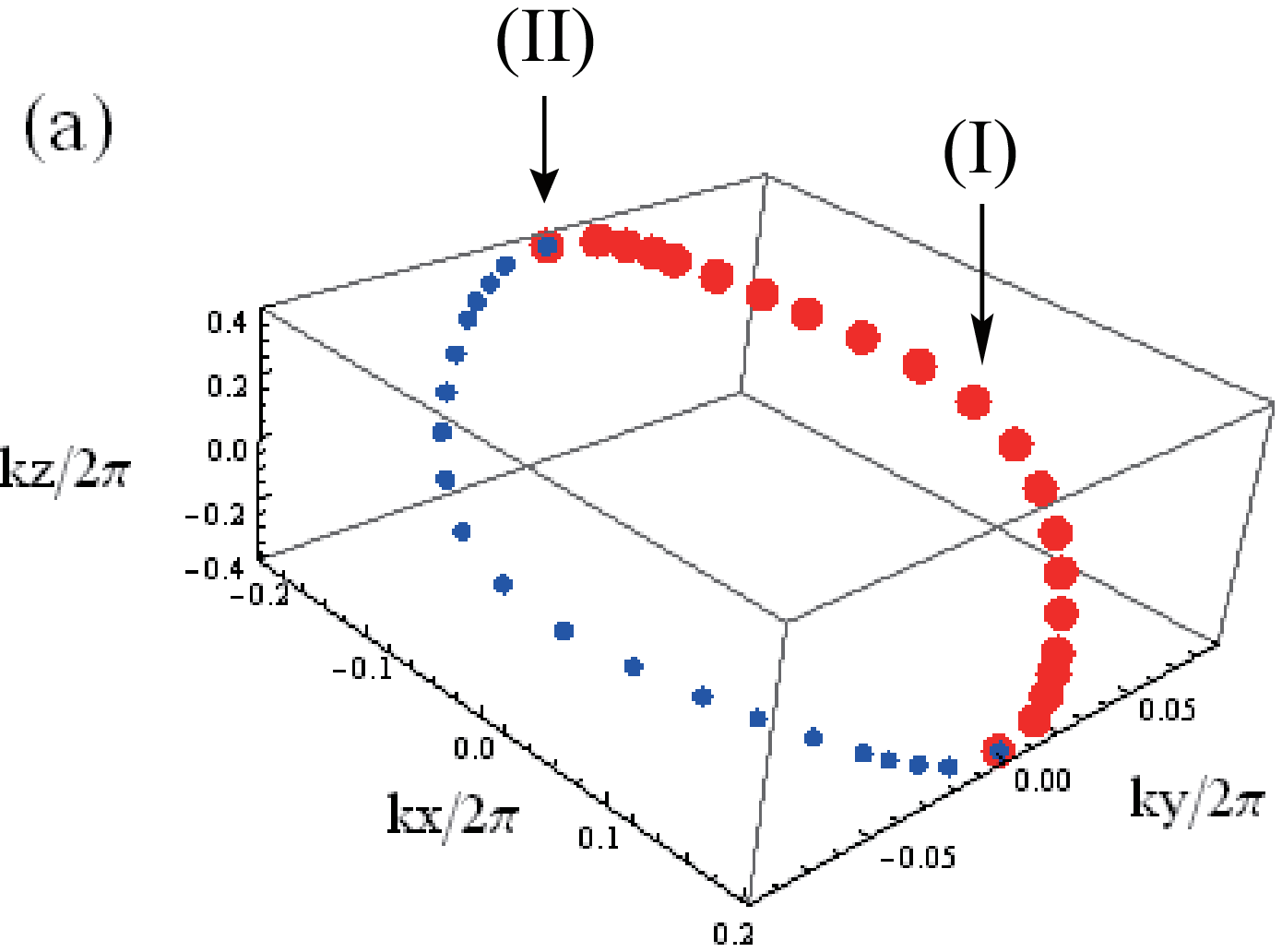}
\includegraphics[width=6cm]{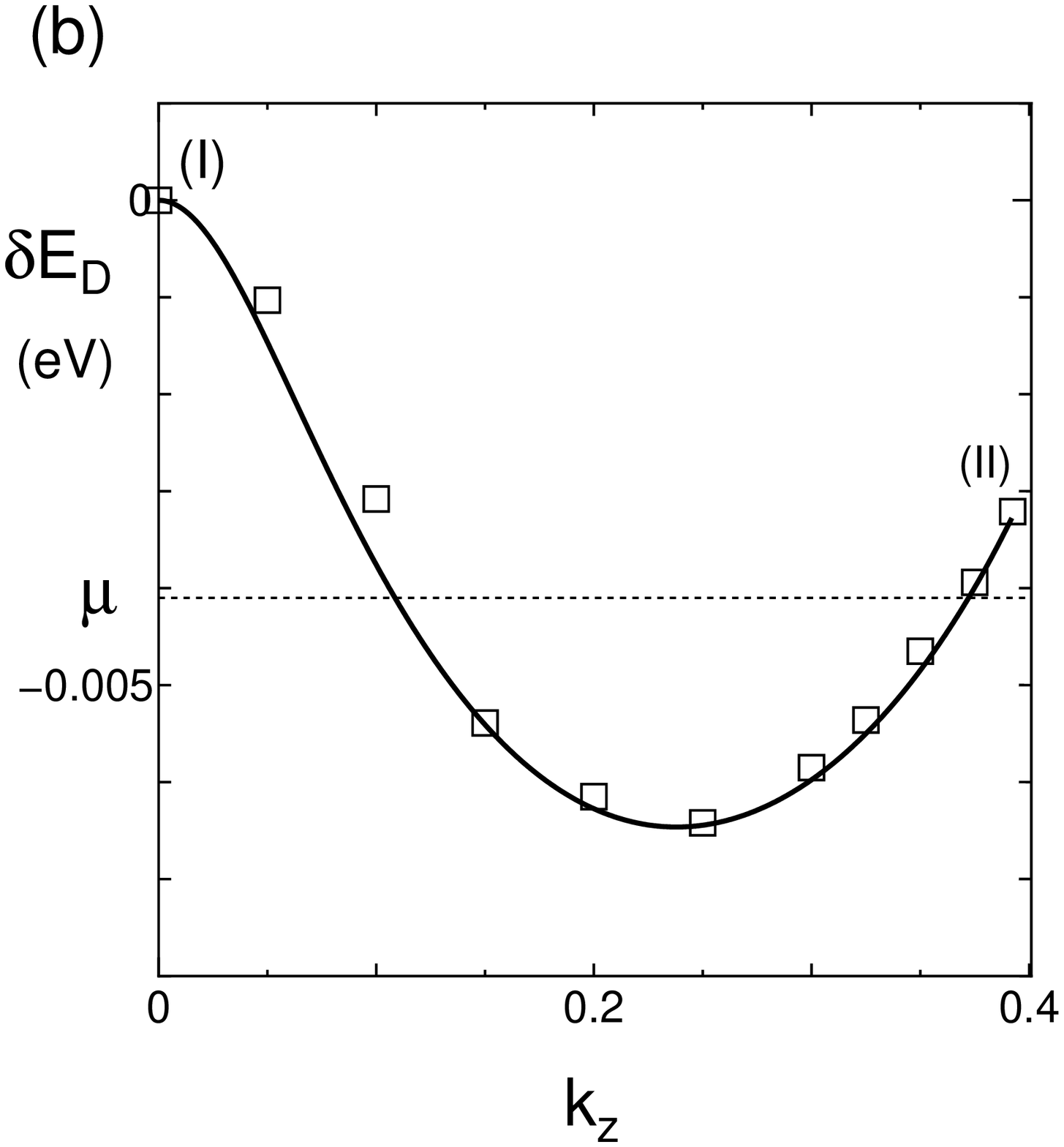}
    \caption{(Color online)
(a) Nodal line in the 3D momentum space $(k_x, k_y, k_z)$, 
 which connects the Dirac point $\bkD [= (\kDx, \kDy, \kDz)]$ 
  calculated from first-principles calculation 
 for the optimized structure at 8 GPa.\cite{Tsumuraya2018_JPSJ} 
For example, Dirac points $\bkD [= (\kDx, \kDy, \kDz)]$ 
which  are scaled by $2\pi$ 
are given by 
(-0.1967, 0.000, 0.3924) (II),
(0, 0.086, 0) (I),
 and 
(0.1967, 0.0, -0.3924) with 
 decreasing $k_z$. The other points are
   taken from  Ref.~\citen{Tsumuraya2018_JPSJ}. 
 These  Dirac points show a mirror  symmetry in the plane of $k_y=0$, i.e.,  
 two points $(\kDx, \pm \kDy, \kDz)$ are symmetric.
The nodal line corresponding to the large symbols (i.e., $k_y \ge 0$ ) 
 is examined in the present paper.
(b) Energy $\delta E_{\rm D} (= E_{\rm D}-C_0)$ in the unit of eV 
along the Dirac point as a 
function  of $k_z (\equiv k_z/2\pi)$,   
  where $C_0$ denotes  $E_{\rm D}$ at  $\bkD/2\pi$ = $(0, 0.086, 0 )$ (I).
For example, $\delta E_{\rm D}$ = 0 for (I),
 -0.00321 for (II), and -0.0064 (minimum) for 
 (-0.143, 0.054, 0.238). 
  The solid line was obtained by substituting $\bkD$ of Eqs.~(\ref{eq:x6a}) and (\ref{eq:x6b}) into Eq.~(\ref{eq:x9}).
The open squares show eigenvalues calculated using the  first-principles band 
 structure. The dotted  line denotes the chemical potential $\mu$ at  
 the Dirac points  
 (-0.085, 0.075, 0.108) and (-0.191, 0.019, 0.372).
 }
\label{fig2}
\end{figure}

 Figure \ref{fig2}(a) shows 
 a nodal line  obtained by 
   the DFT calculation,~\cite{Tsumuraya2018_JPSJ} 
 which is utilized in the present calculation. 
   Although the shape of the line is slightly different 
 from that of the tight-binding model,
the condition of the Dirac point at the TRIM remains the same.~\cite{Tsumuraya_APS}
In the previous paper,~\cite{Tsumuraya2018_JPSJ} 
 it was  shown that 
   the Dirac points  in Fig.~\ref{fig2}(a) can be obtained 
 using  a two-band model of 
  the following effective Hamiltonian $H_{\rm eff}(\bm{k})$
  in the form of a 2 $\times$ 2 matrix;
\begin{eqnarray}
 {H_{\rm eff}(\bm{k})}  
 &=& 
\begin{pmatrix}
f_0(\bk) + f_3(\bk) & - i f_2(\bk)  \\
i f_2 (\bk) & f_0(\bk)- f_3(\bk) 
\end{pmatrix} \ . 
\label{eq:eq1}
\end{eqnarray}
The base  is given by 
 $|H(\bk)>$ and  $|L(\bk)>$, the  wave functions of 
   $H^0(\bk)$  corresponding to HOMO and LUMO, i.e., 
\begin{eqnarray}
  H^0(\bk) |\alpha(\bk)>  =  E_{\alpha}(\bk) |\alpha(\bk)>  \; 
\label{eq:x2}
\end{eqnarray}
 with $\alpha$ = H and L. 
 $\bk = (k_x, k_y, k_z)$ denotes a 3D wave vector. 
$k_x$,  $k_y$, and  $k_z$ correspond to the reciprocal vector 
 for $\bm{a}+\bm{c}$, $\bm{b}$, and $\bm{c}$, respectively.~\cite{Kato_JACS}  
 Matrix elements $f_0(\bm{k})$, $f_2(\bm{k})$, and $f_3(\bm{k})$ 
in Eq.~(\ref{eq:eq1}) are given by 
 \begin{eqnarray}
\label{eq:x3}
 f_2 (\bk)&= & i <{\rm H}(\bk)| H_{\rm int} |{\rm L}(\bk)>\; , 
                                        \\
\label{eq:x4}
 f_3 (\bk)&=&  (E_{\rm H}(\bk) - E_{\rm L}(\bk))/2 \; , 
                                        \\ 
 f_0 (\bk)&=&  (E_{\rm H}(\bk) + E_{\rm L}(\bk))/2 \; , 
\label{eq:x5}
\end{eqnarray}
 where $H_{\rm int}$ denotes the HOMO--LUMO (H--L) interaction. 
Although 
the off-diagonal element is treated by the perturbation, 
such an effective Hamiltonian is justified  
 for the limiting case of 
$f_2(\bk) \rightarrow 0$, which 
 is the present case  of finding the Dirac point.
The energy of Eq.~(\ref{eq:eq1})
 is calculated as  $E_{\pm} =f_0 \pm \sqrt{f_2^2 + f_3^2}$, 
  where  $E_+ (E_-) = E_c (E_v)$ corresponds to the energy of the 
 conduction (valence) band. 
 The Dirac point $\bkD$, which is given by  $E_+=E_-$, 
  is obtained from 
\begin{subequations}
\begin{eqnarray}
   f_2(\bkD) &=& 0 \; , 
\label{eq:x6a}
\\
     f_3(\bkD) &=& 0 \; .
\label{eq:x6b}
\end{eqnarray}
\end{subequations} 
Note that $f_0(\bk)$ and $f_2(\bk)$ are  even functions of $\bk$ 
    because of  time-reversal symmetry  and  
 $f_2(\bk)$  is an odd function of $\bk$ because  the  HOMO and LUMO 
      have different parities.  
Instead of calculating Eq.~(\ref{eq:x2}) directly, 
  we utilize the numerical  results of the DFT calculation as follows.
 The function $f_2(\bk)$ is estimated  
    by projecting the nodal line on the  $k_x$ - $k_z$ plane, 
   while $f_3(\bk)$ is estimated by projecting the nodal line 
     on the $k_x$ - $k_y$ plane.~\cite{Tsumuraya2018_JPSJ} 
 Such a method is justified in the present because of    
    the  presence of  the inversion symmetry  
       at $k_y=0$.

Here we discuss the linear dispersion in
 the present effective Hamiltonian of Eq.~(\ref{eq:eq1}).
Close to the Dirac point,  we rewrite  $f_j(\bk)$ as 
 $f_j(\bk) \simeq f_j(\bkD) +  \bm{v}_j \cdot \delta \bm{k}$  
($j$ = 2, 3 and 0) with $\delta \bk = \bk - \bkD$, 
 where $f_2(\bkD) = f_3(\bkD) =0$ and $f_0(\bkD) \not= 0$. 
 Diagonalizing Eq.~(\ref{eq:eq1}), 
 the energy  of  the Dirac cone is obtained as  
 $E_{\pm}(\bk) \simeq   f_0(\bkD) +  \bm{v}_0 \cdot \delta \bk$ 
 $\pm 
 \sqrt{ (\bm{v}_2 \cdot \delta \bk)^2 + (\bm{v}_3 \cdot \delta \bk)^2}$.
Thus, the energy difference corresponding to  half  the energy difference 
 between the two bands is expressed as
  \begin{eqnarray}
  \Delta_{\bm{k}} &=& [E_{+}(\bm{k})-E_{-}(\bm{k})]/2 
 \nonumber \\
 && = \sqrt{(\bm{v}_2 \cdot \delta \bk)^2 + (\bm{v}_3 \cdot \delta \bk)^2} 
    \; . 
\label{eqx7}
 \end{eqnarray}
Note that the momenta $\delta{\bm k}$ forming this linear dispersion are 
within the  $\bm{v}_2$-$\bm{v}_3$ plane. 
This plane is perpendicular to 
 the tangent of the nodal line since the latter is parallel to  
 $\bm{v}_2 \times \bm{v}_3$.\cite{Suzumura_Yamakage_JPSJ}
 
\subsection{Calculation of  matrix elements}
In this subsection, we examine  $f_2(\bk)$, $f_3(\bk)$, and 
 $f_0(\bk)$ in terms of the  power law  of $\bm{k}$.
Hereafter, we take the lattice constant as unity and 
 scale $k_x$, $k_y$, and  $k_z$ by  2$\pi$, i.e., 
$k_\alpha/2\pi \rightarrow k_{\alpha}$ for $\alpha = x,y$, and $z$.  
The unit of energy is taken as eV.
First,  to reproduce the  nodal line in Fig.~\ref{fig2}(a),~\cite{Tsumuraya2018_JPSJ}
we determine
\begin{subequations}
\begin{eqnarray}
f_2(\bk) & \simeq & C_2(k_z +  k_x + 40 k_x^3 - 380 k_x^5) \; ,
\label{eq:x8a}
     \\
f_3(\bk) & \simeq &  C_3((k_x/0.1967)^2 +  (k_y/0.086)^2 
   \nonumber \\
   & &  +((k_xk_y)^2/0.027^2-1 )
\; .
\label{eq:x8b}
\end{eqnarray}
\end{subequations}
Compared with the previous case,~\cite{Tsumuraya2018_JPSJ} 
  the present calculation was improved by adding 
 arbitrary    $C_2$ and $C_3$. 
    Note that  a non-coplanar nodal line 
 is understood  from  the nonlinear terms in  Eq.~(\ref{eq:x8a}).
 We have determined 
the  coefficients in Eq.~(\ref{eq:x8a}) and (\ref{eq:x8b})
     except for    $C_2$ and $C_3$ 
 by   comparing   
   Eqs.~(\ref{eq:x6a}) and (\ref{eq:x6b}) with 
    Dirac points in Fig.~\ref{fig2}(a).  
 In fact, we used the two Dirac points  
  (0,0.086,0) (I) and  ( - 0.1967, 0, 0.3924) (II), and  
     some  other Dirac points in the intermediate region  in Fig.~\ref{fig2}(a).
Coefficients  $C_2$ and $C_3$,
   which also depend on the location on the nodal line,
     are determined using 
      the velocities of  Dirac points (I) and (II).
Here, the velocity of the cone at the Dirac point
 $\bkD$ is obtained
 as $\bm{v}_2  =\nabla_{\bkD}f_2$,
  $\bm{v}_3  =\nabla_{\bkD}f_3$.~\cite{Suzumura_Yamakage_JPSJ}
 From the DFT calculation, the velocities at  point (I) 
  are  $\vv_2$ = (0.148, 0, 0.148) and $\vv_3$ = (0, 1.25, 0), 
 while 
  the velocities  at  Dirac point (II) 
   are  $v_x \simeq 0.36$ and   $v_z \simeq 0.09$.
  Furthermore, by interpolation between points (I) and (II),  
   we obtain 
  $C_2 = 0.148(1-0.39(k_z/0.392)^2)$
   and  $C_3 = 0.053(1-0.53(k_z/0.392)^2)$ 
  for Eqs.~(\ref{eq:x8a})  and  (\ref{eq:x8b}).

Next, we examine $f_0({\bm k})$ assuming that it has the form
\begin{eqnarray}
f_0(\bk)& \simeq &  b_xk_x^2 + b_yk_y^2 + b_zk_z^2 
   \nonumber \\
  & & + b k_x^2k_z^2 + d_x k_y^2k_x^2 + d_zk_y^2k_z^2 + C_0
   \; ,\label{eq:x9}
\end{eqnarray}
where $f_0(\bkD)$ ( = $E_{\rm D}$) denotes 
 the energy at  Dirac point $\bkD$.

Figure \ref{fig2}(b) shows the energy of the Dirac points 
as a function   of $k_z (\leftarrow k_z/2\pi)$, where  
 the open squares  denote the numerical results of  
   the DFT calculation. 
 Using these data to fit Eq.~(\ref{eq:x9}), we obtain
 $b_x = -0.88, b_y=-2.62, b_z=-0.069$,
 $b=3.7, d_x=-98$, and $d_z=32$.  
 Note that terms with coefficients $b$, $d_x$, and $d_z$ are 
 added in contrast to the previous case~\cite{Tsumuraya2018_JPSJ}, 
    since  terms  with   only $b_x$, $b_y$ and $b_z$ 
     are insufficient to reproduce the data in Fig.~\ref{fig2}(b).
The coefficients $b_x$, $b_y$, $b_z$, and $b$ in 
 $f_0(\bk)$ are determined  from 
  $\delta E_{\rm D} (\equiv   E_{\rm D} - C_0)$ 
 at  Dirac point (II),
 and the tilting velocities 
 $\bm{v}_0 = (0, -0.45, 0)$ and $\bm{v}_0 = (0.12, 0, 0.06)$ 
     at  Dirac points (I) and (II), respectively.  
$C_0$ denotes  $E_{\rm D}$ at  Dirac point (I). 
Furthermore,  coefficients $d_x$ and $d_z$ are determined 
  from Dirac points (symbols) with 
 $\delta E_{\rm D}$ = -0.0539 and -0.0642 close to the minimum
   in Fig.~\ref{fig2}(b). 
 The energy $E_{\rm D}$ (solid line) 
 is calculated by substituting  
the Dirac point into Eq.~(\ref{eq:x9}), where $\bkD$ is obtained from  
 Eqs.~(\ref{eq:x6a}) and (\ref{eq:x6b}). 
It turns out that   $E_{\rm D}$ (solid line) coincides reasonably well with 
  that obtained from first-principles calculation (open squares).

 The chemical potential $\mu$ (dotted line)  
  is obtained from the condition of the half-filled band, which is shown later.  It is found that 
 the Fermi surface cuts the entire line eight times followed by 
 the alternation of  the hole and electron pockets, e.g., 
 the hole pockets are obtained for (I) and (II).

From Eqs.~(\ref{eq:x8a}), (\ref{eq:x8b}), and (\ref{eq:x9}), 
 the explicit form of 
  $\bm{v}_j$ is  given as 
\begin{subequations}
\begin{eqnarray}
\label{eq:x10a}
 \bm{v}_2 & = & \nabla_{\bkD}f_2  \simeq  
         C_2(1 + 120 \kDx^2 - 1900\kDx^4, 0, 1), 
                                          \\
\label{eq:x10b}
 \bm{v}_3 & = & \nabla_{\bkD}f_3   \simeq  
       C_3(2\kDx/0.1967^2+2\kDx\kDy^2/0.027^2, 
                                     \nonumber \\
      & &2\kDy/0.086^2 + 2\kDx^2\kDy/0.027^2, 0)
                               \; ,  \label{eq:x10b} \\
\label{eq:x10c}
 \bm{v}_0 & = & \nabla_{\bkD}f_0 \simeq (
 2\kDx (b_x + b  \kDz^2  + d_x \kDy^2),
   \nonumber \\
& & 2\kDy (b_y + d_x \kDx^2 + d_z \kDz^2), 
 \nonumber \\
& & 2\kDz (b_z + b  \kDx^2 + d_z \kDy^2)
)   
\; .
   \label{eq:x10c} 
 \end{eqnarray}
\end{subequations}
Although the derivatives of $C_2(\bkD)$ and $C_3(\bkD)$ with respect to 
 $\bkD$ are finite,   
  Eqs.~(\ref{eq:x10a})  and  (\ref{eq:x10b}) are still valid 
    owing to   $f_2(\bkD)/C_2 =0$ and   $f_3(\bkD)/C_3\ =0$.

 Here, we mention the behaviors of the velocity of the Dirac cone 
    in the region of $-0.3924 \le k_z \le 0.3924$ and
  for $k_y \ge 0$,  which corresponds to the line 
          given by the large symbols in Fig.~\ref{fig2}(a).
 An arbitrary  $\bkD$ is  calculated 
 self-consistently  from 
      Eqs.~(\ref{eq:x6a}) and (\ref{eq:x6b}) 
       with Eqs.~(\ref{eq:x8a}) and (\ref{eq:x8b}).
Using these Dirac points, 
    the velocities of the Dirac cone $\bm{v}_2$ and $\bm{v}_3$  
        are obtained from Eqs.~(\ref{eq:x10a}) and (\ref{eq:x10b}). 
Velocities  $\bm{v}_2(\bkD)$ and $\bm{v}_3(\bkD)$    
  as a function of  $k_z$  show that  
 $v_{2x}$, $v_{2z}$, and $v_{3y}$ are even  but $v_{3x}$ is  odd.
The tilting velocities of the cone $\bm{v}_0(\bkD)$ 
 as a function of  $k_z$ show that 
 $v_{0y}$ and $v_{0x}$ are  even 
  but   $v_{0z}$ is odd. 
 These properties originate 
   from  $f_3(\bm{k})$ and $f_0(\bm{k})$ 
    being  even and  $f_2(\bm{k})$ being  odd 
   with respect to $\bm{k} \rightarrow - \bm{k}$. 

\section{Properties of Dirac Cone}

\subsection{Unit vector along nodal line}
Since  $\bm{v}_2$  is not orthogonal to $\bm{v_3}$ except for $k_{0z} = 0$,
 we calculate the  principal axes to understand  clearly 
 the Dirac cone for an  arbitrary Dirac point on the nodal line. 
  First,  
we  introduce a set of three orthogonal unit vectors, 
  $\ve_1$, $\ve_2$,  and $\ve_{\perp}$.
 Quantities     $\ve_2$, $\ve_3$, and $\ve_1$  are  unit vectors parallel to 
    $\vv_2$, $\vv_3$, and $\vv_2 \times \vv_3$, respectively. 
Since the direction of $\ve_1$ is  the tangent of the nodal line, 
 the vectors of principal axes for the Dirac cone are located 
 on the plane perpendicular to $\ve_1$, i.e., 
 on the $\ve_2$-$\ve_3$ plane. 
 To consider the orthogonal basis on the $\ve_2$-$\ve_3$ plane, 
  we introduce    $\ve_{\perp} ( = \ve_1 \times \ve_2)$, which is  orthogonal 
    to both $\ve_1$ and $\ve_2$. 
 These vectors  expressed as 
\begin{subequations}
\begin{eqnarray}
\label{eq:x10a}
& & \ve_2 = \vv_2/v_2  = (v_{2x},0, v_{2z})/v_2 \; ,
  \\
& & \ve_3 = \vv_3/v_3  = (v_{3x}, v_{3y}, 0)/v_3 , \\
\label{eq:x11b}
& & \ve_1= \ve_2 \times \ve_3/|\ve_2 \times \ve_3| \; , \\
\label{eq:x11c}
& & \ve_{\perp} = \ve_1 \times \ve_2 \; , 
 \end{eqnarray}
\end{subequations}
 where $v_2 = \sqrt{v_{2x}^2+v_{2z}^2}$ and 
       $v_3 = \sqrt{v_{3x}^2 + v_{3y}^2}$. 
 Figure \ref{fig3}(a) shows 
 the components of $\ve_1$ as a function of $k_z$; 
 $e_{1y}$ is odd  while  $e_{1x}$ and $e_{1z}$ are even.  
With increasing $k_z$,  $e_{1y}$ changes  from 1 to -1, while 
 the signs of $e_{1x}$ and $e_{1z}$ remain unchanged.  
Note that  
 $\ve_1$ with  $k_y < 0$ [small symbols in Fig.~\ref{fig2}(a)]  
is obtained 
 from  $\ve_1$ with  $k_y > 0$
 by the replacement 
  $(\kDx, \kDy, \kDz) \rightarrow (-\kDx, \kDy, -\kDz)$.

\subsection{Principal axes and velocities}
Next, we examine the principal axes of the linear dispersion $\Delta_{\bm{k}}$  
  of Eq.~(\ref{eqx7}), which is expressed  
    in terms of $\ve_2$ and $\ve_{\perp}$.
 Since    
 $\bm{v}_2$ is not orthogonal to $\bm{v}_3$  except when $k_z = 0$,
 we introduce $\phi$ as the  angle between $\vv_2$ and $\vv_3$,
\begin{subequations}
\begin{eqnarray}
\cos \phi = (\bm{v}_2 \cdot \bm{v}_3)/(v_2v_3)\; ,  
 \label{eq:x12a}
\end{eqnarray}
where 
$\phi - \pi/2$ is an odd function of $k_z$ 
 and $|\cos \phi|$ increases monotonically with  $|k_z|$. 
 When $\Delta_{\bk}$ is expressed in terms of the principal axes, we note that 
 $\Delta({\bq})$, where   $\vq = \vk - \vk_0  = q_1\ve_1 + q_2 \ve_2 + q_3\ve_{\perp}$, is written in terms of the principal axes.   
 Noting that  $\cos \phi = \ve_2 \cdot \ve_3$, 
   $\vv_2 = v_2 \ve_2$, and $\vv_3 = v_3 \cos (\phi) \ve_2 +v_3 \sin (\phi) \ve_{\perp}$, we obtain   
\begin{eqnarray}
\label{eq:x12b}
& &  \vv_2 \cdot \vq = v_2q_2 \; ,  \\
\label{eq:x12c}
& & \vv_3 \cdot \vq = v_3 q_2 \cos \phi + v_3 q_3 \sin \phi \; . 
 \end{eqnarray}
\end{subequations}
Thus, the explicit form of  
 $  \Delta (\bm{q})$ of Eq.~(\ref{eqx7}) is written as   
\begin{subequations}
\begin{eqnarray}
\label{eq:x12a}
 \Delta (\bm{q})^2 &=&  (\vv_2 \cdot \vq)^2 + (\vv_3 \cdot \vq)^2  \nonumber \\
& & =  Aq_2^2 + 2 C q_2q_3 + B q_3^2 \; ,  \\
\label{eq:x12b}
 A & = & v_2^2+(v_3 \cos \phi)^2  \; , \\
\label{eq:x12c}
B & = & v_3^2 (\sin \phi)^2 \; ,\\
\label{eq:x12d}
 C & = &  v_3^2 \sin \phi \cos \phi\; . 
 \end{eqnarray}
\end{subequations}
 The principal axes are obtained 
 by  rotation   from 
  the $q_2 - q_3$  plane to  the $q_- - q_+$  plane  
 to eliminate the second term that is proportional to $q_-q_+$. 
The result is obtained as    
\begin{subequations}
\label{eq:delta}
\begin{eqnarray}
\label{eq:x14a}
 \Delta (\bm{q})^2
 &= & V_+^2 q_+^2 + V_-^2q_-^2 \; ,\\
\label{eq:x14b}
V_{\pm}^2& =& \frac{1}{2}\left[A + B \pm \sqrt{(A-B)^2+ 4C^2} \right] \; ,\\
\label{eq:x14c}
 \tan (2 \theta) &=& \frac{2C}{A - B}
 = \frac{v_3^2 \sin 2 \phi}{v_2^2 + v_3^2 \cos 2 \phi}
\; ,
 \end{eqnarray}
\end{subequations}
where $q_+$ and $q_-$ are the rotated coordinates of principal axes given by
\begin{subequations}
\begin{eqnarray}
\label{eq:x15a}
& &  \vq = q_1 \ve_1+ q_+\ve_{+} + q_- \ve_{-} \; , \\
\label{eq:x15b}
 & & \ve_{-} = \cos \theta  \ve_2 + \sin \theta \ve_{\perp} \; , \\
\label{eq:x15c}
 & & \ve_{+} = - \sin \theta \ve_2 + \cos \theta \ve_{\perp} \; .
 \end{eqnarray}
\end{subequations}
  $\theta$ is the angle  between $\ve_{-}$ and $\ve_2$ and  
 is chosen to be   $|\theta| \leq \pi/2$. 
 $V_+$ and $V_-$ are the velocities of the principal axes. Note that  
$V_+ > V_-$.

\begin{figure}
  \centering
\includegraphics[width=4cm]{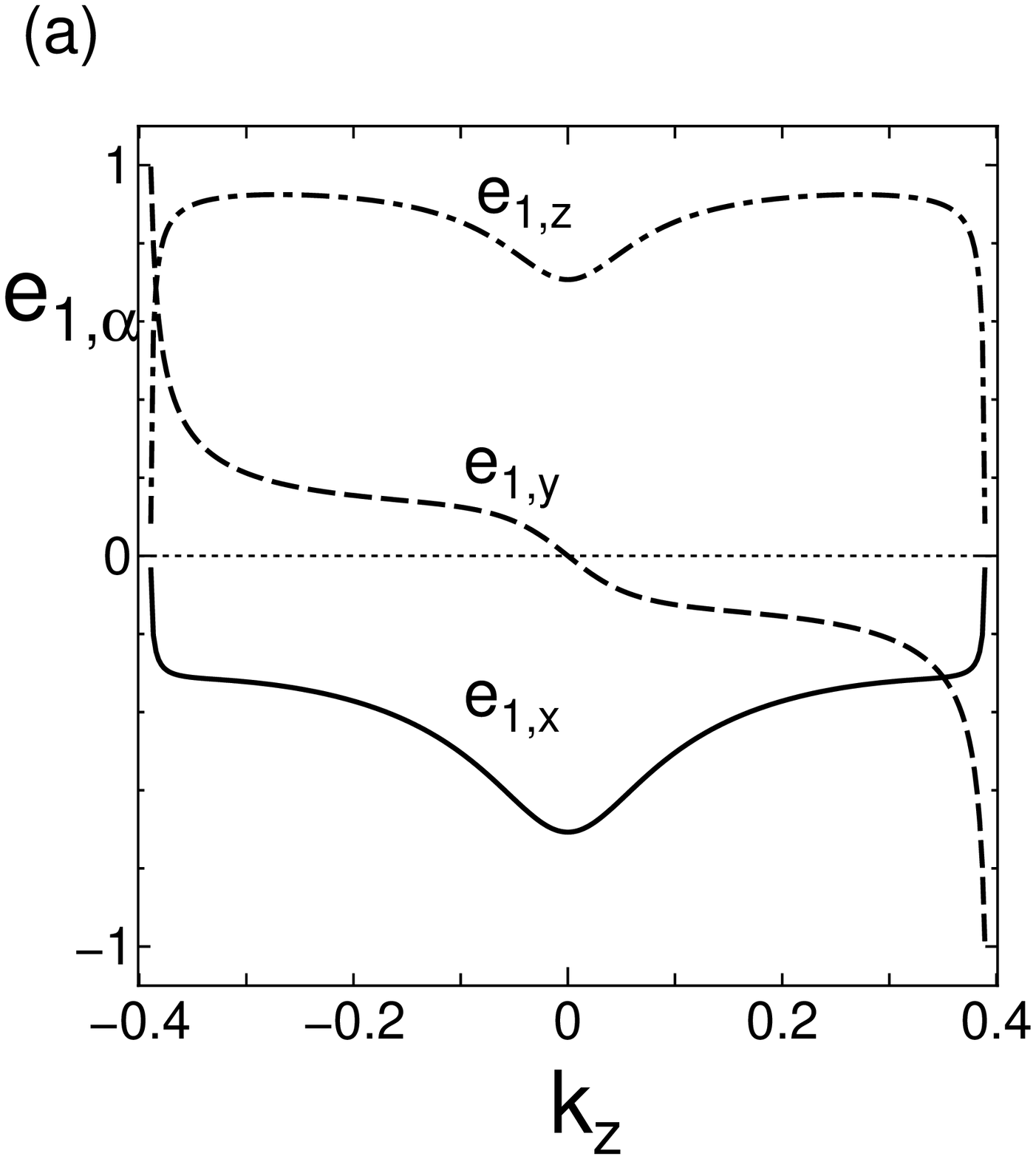} 
\includegraphics[width=4cm]{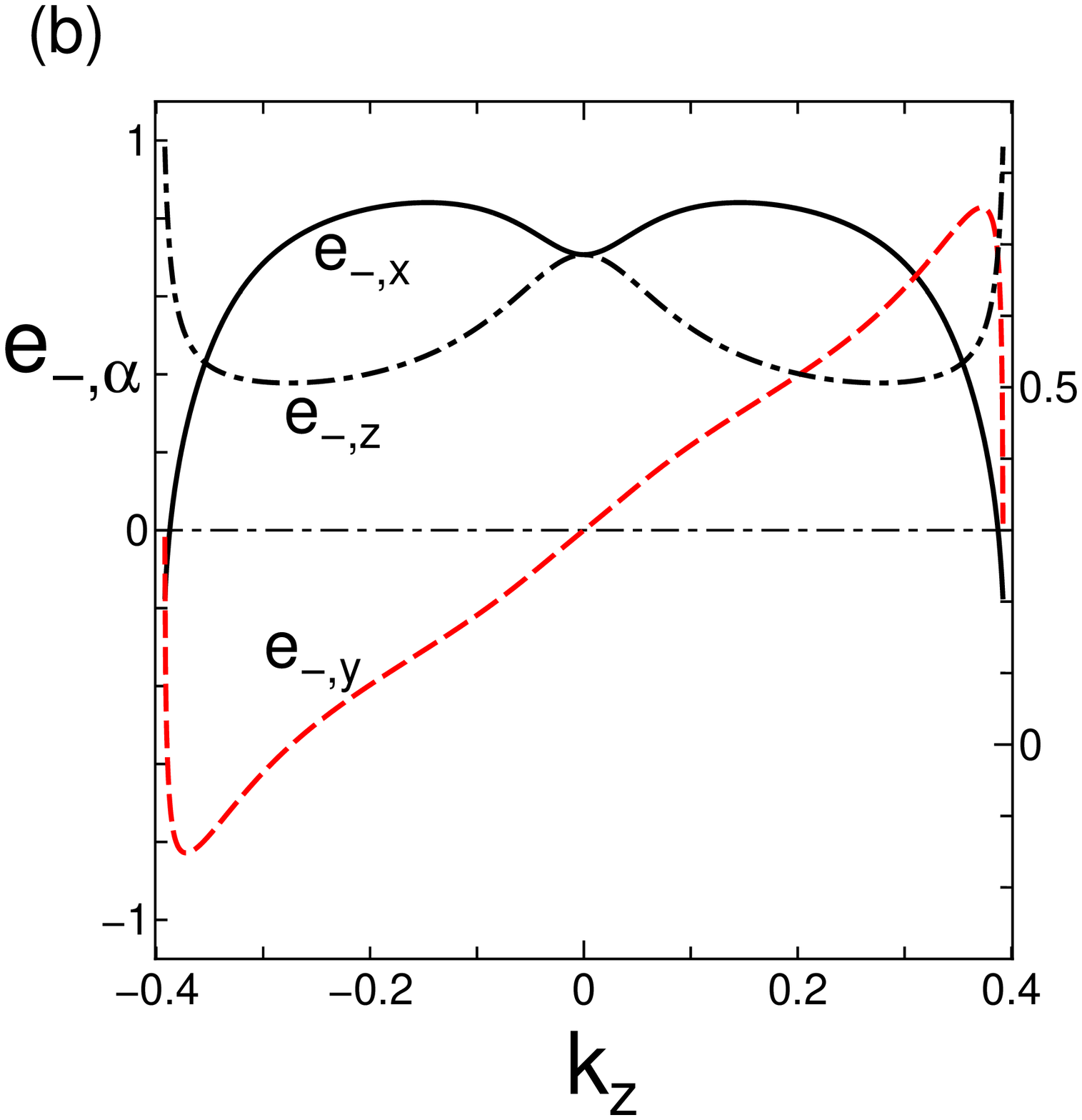} 
\includegraphics[width=4cm]{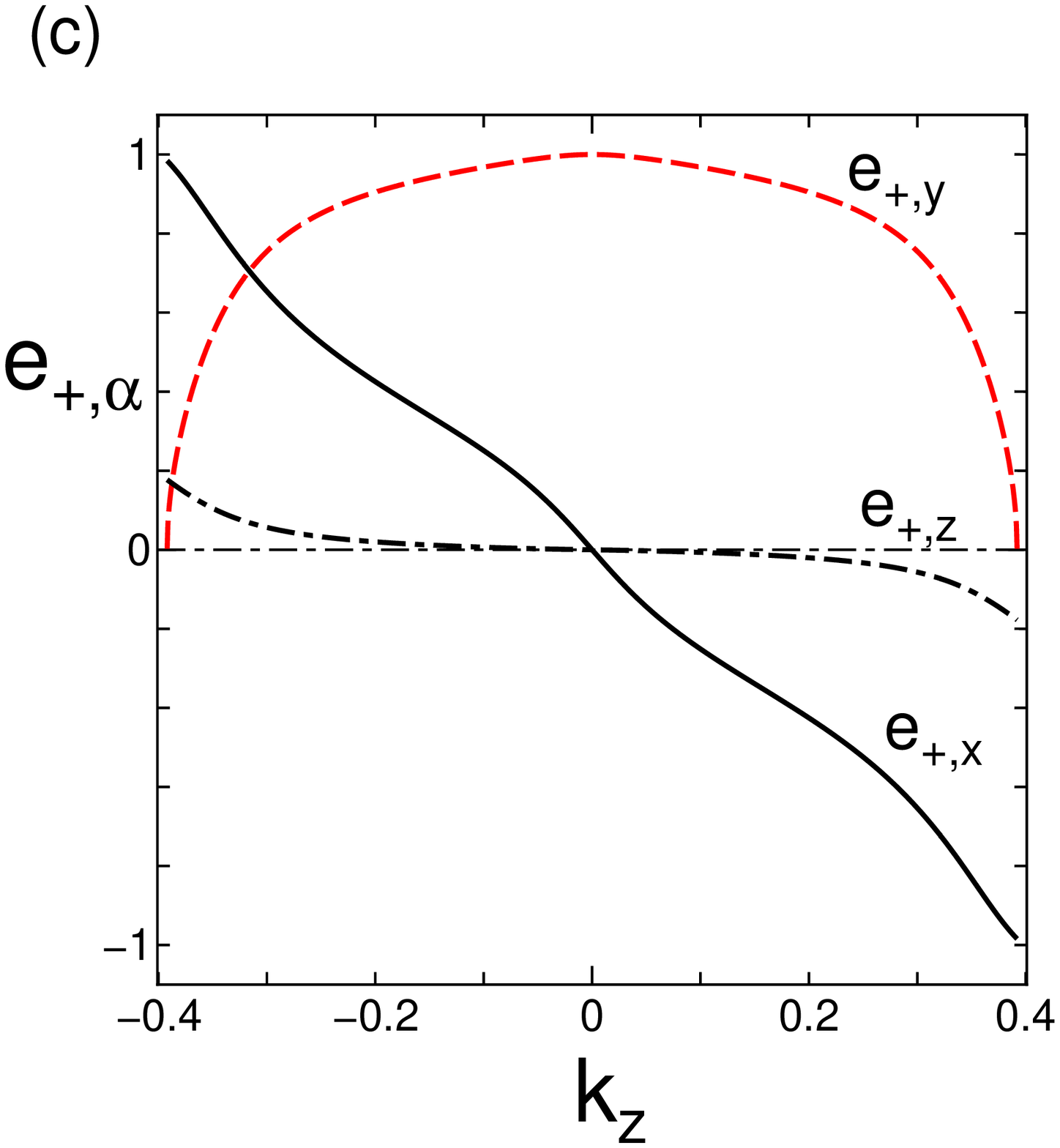} 
    \caption{(Color online)
(a) $k_z$ dependence  of unit vector $\ve_{1}$ [Eq.~(\ref{eq:x11c})] 
 parallel to the nodal line, where the  components are given  by 
  $e_{1,x}$, $e_{1,y}$, and   $e_{1,z}$.  
 $e_{1y}$ is  odd while 
  $e_{1x}$ and  $e_{1z}$ are even. 
$\ve_1 = (-1, 0, 1)/\sqrt{2}$ for $k_z = 0$
 and $\ve_1 = (0, \mp 1, 0)$ for $k_z = \pm 0.3924$. 
(b) $k_z$ dependence  of unit vector $\ve_{-}$ [Eq.~(\ref{eq:x15b})] 
for the principal axis 
 of $V_-$, where the  components are given  by 
  $e_{-,x}$, $e_{-,y}$, and   $e_{-,z}$.  
(c) $k_z$ dependence  of unit vector  $\ve_{+}$, [Eq.~(\ref{eq:x15c})], 
for the principal axis  of $V_+$,
 where the components 
  are given by  $e_{+,\alpha}$ ($\alpha = x, y, z)$.
 The vector  $\ve_{+}$ is also obtained from $\ve_+ = \ve_1 \times \ve_{-}$
  with  $\ve_1$ in (a).  
 }
\label{fig3}
\end{figure}

Figures \ref{fig3}(b) and \ref{fig3}(c) show 
   the   $k_z$ dependence of the unit vector, 
       $\ve_\pm$, for the respective  principal axes.
Figure \ref{fig3}(b) shows  the 
  component of $\ve_- =(e_{-,x},e_{-,y},e_{-,z})$. 
 As a function of $k_z$, $e_{-,x}$ and $e_{-,z}(>0)$  are even and  
  $e_{-,y}$ is odd. 
    $e_{-,z}$ takes a minimum and 
       $ \simeq 1$ for $k_z  = \pm 0.3924$, 
        while $e_{-,x}$ takes a maximum and 
     decreases almost to zero  for $|k_z| \simeq 0.3924$. 
 With increasing $|k_z|$,   $|e_{-,y}|$ increases linearly, followed by 
 a sudden decrease to zero at $|k_z|  \simeq  0.3924$.
Figure \ref{fig3}(c) shows the  $k_z$ dependence of the  
   component for  $\ve_+ =(e_{+,x},e_{+,y},e_{+,z})$. 
 $e_{+,y}$ is an even function, where  $e_{+,y} = 1$ at $k_z =0$ and decreases 
     to zero monotonically  with $|k_z|$ increasing to 0.3924. 
 $e_{+,x}$ and $e_{+z}$ are odd functions, 
   and $e_{+,x}$ changes from $\simeq 1$ 
  to $\simeq -1$. 
  The variation of  $e_{+,z}$ is much smaller than that of $e_{+,x}$. 
  The rotation of $\ve_+$ with $k_z$ increasing from $0$ to $0.3924$ 
 is also reasonable compared with 
 that of $\ve_-$,  because  $\ve_+ \cdot \ve_- = 0$.

The cross section for $\Delta(\bk) = E_0 $  
 is an ellipse   with the radius of the  minor (major) axis 
  calculated as  $a=E_0/V_+$ ($b=E_0/V_-$).
Using  $V_+$ and $V_-$, 
  the area of the ellipse, $S$, for  the gap 
      $2E_0 = E_+ - E_-$  
 is given by  
 $S(\bkD) = \pi ab = \pi E_0^2/(V_+V_-)$ 
 $= \pi E_0^2/ \sqrt{AB - C^2}$ 
$ = \pi E_0^2 / |\bm{v}_2(\bkD) \times \bm{v}_3(\bkD)|
 = \pi E_0^2/(v_2v_3 |\sin \phi|)$.
As a function of $|k_z|$,
$S$ is almost constant but exhibits  a rapid increase at  
 large $|k_z|$.

\begin{figure}
  \centering
\includegraphics[width=6cm]{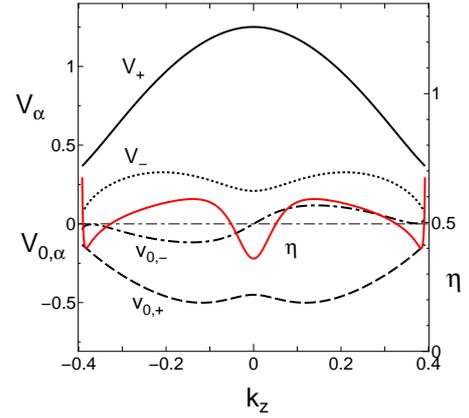} \\
    \caption{(Color online)
 Velocities  $V_+$ and $V_-$,   
   obtained from  Eq.~(\ref{eq:x14b}). 
$v_{0\pm}$  [Eq.~(\ref{eq:x16b})], denotes the tilting velocity for the corresponding axis. 
 The quantity $\eta$ denotes a tilting parameter given by 
 $\eta = ((v_{0,-}/V_-)^2 + (v_{0,+}/V_+)^2)^{1/2}$, 
 as shown in Eq.~(\ref{eq:x18c}). 
 }
\label{fig4}
\end{figure}


Figure \ref{fig4} shows the velocities  $V_{\pm}$ obtained from 
  Eq.~(\ref{eq:x14b}).  
 The principal axes of the ellipsoid 
    are obtained by a rotation of  $\theta$ 
        from the $q_x$-$q_y$ plane to the $q_-$-$q_+$ plane.
The quantity $\theta$ is odd with respect to $k_z$ and becomes $\simeq \pm \pi/2$  
      for $k_z= \pm 0.3924$. 
The velocity  $V_+$ decreases monotonically but $V_-$ takes a maximum 
 with increasing $|k_z|$.  
 Principal values $V_{\pm}$  ($V_+ > V_-$) show  
   a large  anisotropy, where 
  $V_+/V_-$ is  maximum ($\simeq$ 6.0) at  $k_z = 0$ 
 and  minimum ($\simeq$ 2.3) at $|k_z|$ = 0.324. 

\subsection{Effect of tilting}
We briefly mention  the Dirac cone in the presence of 
  the tilting velocity $\vv_0$. 
In terms of  $\ve_{\pm}$, the tilting velocity is rewritten   as  
\begin{subequations}
\begin{eqnarray}
\label{eq:x16a}
\vv_0 &= &  (v_{0x}, v_{0y}, v_{0z}), \nonumber \\
 & = &  v_{0,1} \ve_1  + v_{0,-}\ve_{-} + v_{0,+}\ve_{+} \; , \\
 v_{0,\pm}&=& \vv_0 \cdot \ve_{\pm} \; ,
\label{eq:x16b}
 \end{eqnarray}
\end{subequations}
 with $v_{0,1} = \vv_0 \cdot \ve_1$.
Figure \ref{fig4} shows  $v_{0,\pm}$, where 
 $v_{0,-}$  ( $v_{0,+}$) is an odd (even) function with respect to 
$k_z$.
By taking account of  $\vv_0 \cdot \bm{q}$ 
 with $\bm{q} = q_-\ve_{-} + q_+ \ve_{+}$, 
 the energy of the upper band $E_{+}(\bm{q}) = E$  
 is written as 
\begin{eqnarray}
\label{eq:x17}
\sqrt{(V_-q_-)^2 + (V_+q_+)^2} + v_{0.-}q_{-}+v_{0,+}q_{+} = E \; .
 \end{eqnarray}

 Defining $\tq_{\pm} = V_{\pm}q_{\pm}$,
 we examine the tilting on the plane 
 of $\tq_-$ and $\tq_+$.
Equation (\ref{eq:x17}) is rewritten as 
 \begin{subequations}
\begin{eqnarray}
\label{eq:x18a}
  & & \sqrt{\tq_-^2+ \tq_+^2} +\td \cdot (\tq_-\ve_- + \tq_+\ve+)= E \; , \\
\label{eq:x18b}
 & & 
  \td = (v_{0,-}/V_-) \ve_{-} + (v_{0.+}/V_+) \ve_{+} \; ,
\\
\label{eq:x18c}
  & & \eta = |\td| = \sqrt{ (v_{0,-}/V_-)^2 +(v_{0,+}/V_+)^2 } \; .
\end{eqnarray}
 \end{subequations}
 The quantity  $\eta$ denotes a tilting parameter and  
  its $k_z$ dependence is shown in Fig.~\ref{fig4}.
 The Dirac cone is tilted but not overtilted 
 because   $\eta < 1$.
Defining $\theta'$ by   $\ve_- \sin \theta' - \ve_+ \cos \theta' = \td/|\td|$, 
 Eq.~(\ref{eq:x18a}) is  rewritten  as 
 \begin{subequations}
\begin{eqnarray}
& &
\label{eq:eq19a}
(1 - \eta^2) Q_-^2 
   + (1-\eta^2)^2 \left( Q_+ - \frac{E\eta}{1- \eta^2} 
    \right)^2 = E^2 \; , \;\;\; \nonumber \\
\end{eqnarray}
 where  
\begin{eqnarray}
\label{eq:eq19b}
& &  \tq_- \ve_{-} + \tq_{+} \ve_{+} 
   = Q_- \ve_{\delta 2} + Q_{+} \ve_{\delta 1} \; , 
           \\ 
\label{eq:eq19c}
  & &  \ve_{\delta 2} = \cos \theta' \ve_- + \sin \theta' \ve_+ \; ,
 \\
\label{eq:eq19d}
 & &  \ve_{\delta 1} = - \sin \theta' \ve_- + \cos \theta' \ve_{+} \; ,
 \\
\label{eq:eq19e}
 & &  \sin  \theta' =  \frac{(v_{0,-}/V_-)}{\eta}  \; .
 \end{eqnarray}
\end{subequations}
Equation (\ref{eq:eq19a}) shows 
   an ellipsoid with  radius  
     $E(1 - \eta^2)^{-1}$ [$E(1-\eta^2)^{-1/2}$] for $Q_+ (Q_-)$. 
The center is located at 
        $[E\eta/(1 - \eta^2)](-\sin \theta', \cos \theta')$ 
 on the plane  of $\tq_-$ and $\tq_+$.
The phase $\theta'$ is the angle between $\ve_{\delta,2}$ and $\ve_{-}$, where 
$\ve_{\delta, 1}$ and $\ve_{\delta, 2}$ 
 are orthogonal to each since due to $\ve_{+} \cdot \ve_- = 0$.
 For $k_z=0$, the principal axis   
 is given by $\ve_+=  \ve_{\delta 1}$, i.e., 
 $\tq_+ = Q_{+}$    
   with  $\theta' = 0$ owing  to $v_{0,-}=0$.  
 Thus, the rotation angle of the axes of the ellipsoid 
 is  obtained as $\theta'$, i.e.,   
$Q_- = \tq_- \cos \theta' + \tq_+ \sin \theta'$  and 
$Q_+ = - \tq_- \sin \theta' + \tq_+ \cos \theta' $
from Eqs.~(\ref{eq:eq19b}), (\ref{eq:eq19c}), and (\ref{eq:eq19d}).
Note that it is straightforward to 
 calculate the  anisotropic conductivity  by 
 projecting the  electric field on the axes of 
$\ve_{\delta, 1}$ and $\ve_{\delta, 2}$.~\cite{Suzumura2014}  

\begin{figure}
  \centering
\includegraphics[width=6cm]{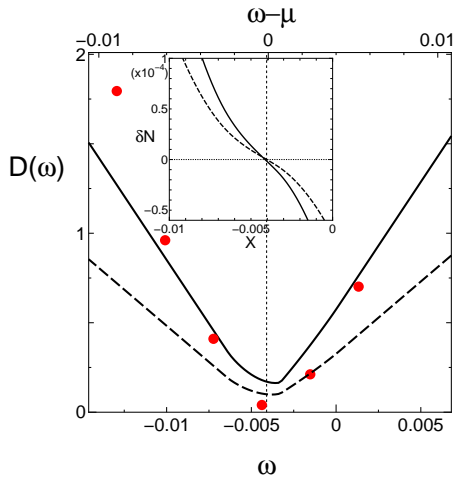}
    \caption{(Color online)
Density of states (DOS) as a function of $\omega$
 (solid line),  which is given by Eq.~(\ref{eq:x21}).
 The origin is taken at $E_d$ of  Dirac point (1). 
 The dashed line denotes the DOS without tilting. 
The inset denotes the corresponding $\delta N$, which is given by 
 Eq.~(\ref{eq:x20}).
The vertical dotted  line denotes the location of the chemical potential.
The closed circles  denote DOS values obtained by 
  the first-principles DFT calculation. 
 }
\label{fig5}
\end{figure}

\section{Electronic States and Response to Magnetic Field}
In this section, the present effective Hamiltonian is applied to 
 calculate the density of states and orbital magnetic susceptibility. 

\subsection{Density of states}

To calculate the number of states, we note that 
the area of the ellipse of the Dirac cone  with $ \mu =  E_0$ is given by 
$S(\bkD) =  \pi E_0^2/(V_+V_-)$. Here $V_+V_-$ turns out to be $v_2v_3 |\sin \phi|$
from Eqs.~(\ref{eq:x12b}), (\ref{eq:x12c}), (\ref{eq:x12d}),
  (\ref{eq:x14a}), and (\ref{eq:x14b}).
   Furthermore, this area is modified as 
$S/(1 -\eta^2)^{3/2}$ in the presence of tilting, as can be seen from Eq.~(\ref{eq:eq19a}).
Taking the origin of the number at the respective  Dirac point,
 the deviation of the total number of states 
 from  that of  a half-filled band
is calculated as
\begin{eqnarray}
\delta N(X) &=& \int_{\overline{C}} {\rm d} \bm{s} \cdot \ve_1 \;\;\;
  (\delta E_D(\bm{k}_0) - X)^2 \nonumber  \\
\label{eq:x20}
    & & \times 
 \frac{2 \pi \; {\rm sgn} (\delta E_D(\bm{k}_0) - X)}{v_2v_3|\sin (\phi)| (1 -\eta^2)^{3/2}}\; , 
\end{eqnarray}
where $X$ and $\delta E_D(\bm {k}_0)$ are 
measured from the energy of the Dirac point (I), i.e., 
$E_D(\bm{k})$ at $\bm{k} =(0, 0.086,0)$. 
 The quantity $X$ is introduced as a  chemical potential, which 
 gives the variation of $\delta N(X)$.  
Here, 
 $ \int_{\overline{C}}$ represents the integral 
  [corresponding to large symbols in Fig.~\ref{fig2}(a)], 
 and    the integral is performed using 
      $\int {\rm d} \bm{s} \cdot \ve_1 = \int {\rm d} z /\e_{1z}$.  
From $\delta N(X)$, the DOS is given by
\begin{eqnarray}
\label{eq:x21}
D(\omega)&= & - \frac{\partial (\delta N)}{ \partial X}\vert_{X=\omega}
  \; .
 \end{eqnarray}

In  Fig.~\ref{fig5}, the solid lines show $D(\omega)$ and $\delta N(X)$ (in the inset) as a 
function of $\omega$ (or $X$). 
Note that $X$ at which $\delta N(X) = 0$ holds corresponds to the 
chemical potential $\mu$. This gives $\mu = -0.0041$, as can be seen in Fig.~\ref{fig5}.
The dashed line denotes the DOS without tilting, i.e., $\eta = 0$, which is lower 
than the solid line, because the tilting increases the area 
 of the ellipsoid, $S$, with  fixed energy $E_0$.
 Within the numerical accuracy, 
 one finds the relation (see the top $x$-axis in  Fig.~\ref{fig5}) 
 \begin{eqnarray}
\label{eq:x22}
 D(\omega - \mu) \propto |\mu -\omega|  \; , 
 \end{eqnarray}
for $ 0.005< |\omega -\mu| $, 
 while there is a slight deviation  
 for $|\omega -\mu| < 0.005$. 
 This comes from the nonmonotonic variation of $E_D$ with respect to $k_z$, 
 as seen in Fig.~\ref{fig2}(b).

Next, we compare the above results with those in the DFT calculation. 
 To obtain DOS using the first-principles DFT method, 
 we need a more elaborate numerical
calculation than that described  in Sect. 2. 
For this purpose,
$\bf{k}$-point meshes are taken as 16 $\times$  32 $\times$  16 
 for the DFT-optimized structure under the pressure of 8 GPa,~\cite{Kato_JACS, Tsumuraya2018_JPSJ}
where Kohn--Sham equations are self-consistently solved in a scalar-relativistic fashion by the all-electron full-potential linearized augmented plane
wave (FLAPW) method\cite{Wimmer_FLAPW81,KA_FLAPW81,Weinert81} within an exchange-correlation functional of a generalized gradient approximation (GGA)\cite{PBE}. The obtained DOS values are shown in Fig.~\ref{fig5} by the closed circles. 
The finite DOS for small $\omega$ suggests  
  a metallic behavior. This is qualitatively consistent 
with the experimental results~\cite{Suzumura2018_JPSJ_T}
 and also with the solid line in Fig.~\ref{fig5}, 
 while the behavior  close to the  minimum 
shows  a deviation from the solid line. 
Thus, the  present calculation in terms of the  effective model 
 may provide  reasonable  results for a nodal line semimetal. 
\vspace{1cm}

\subsection{Some properties  of non-coplanar nodal loop}

In this subsection, we clarify several properties of the nodal line (loop). 
First, we define the average for some quantity $F$ 
 along the nodal line by 
\begin{eqnarray}
 < F > =  \frac{\int_{\overline{C}} {\rm d} \bm{s} \cdot \ve_1 \; F }{
   \int_{\overline{C}} {\rm d} \bm{s} \cdot \ve_1 } \; .  
\label{eq:x23}
\end{eqnarray}
In the present nodal line, a length of the line is  obtained as 
$L = 2  \int_{\overline{C}} {\rm d} \bm{s} \cdot \ve_1$ = 1.856, 
which  is smaller than 4, being 
 the length of the square of the Brillouin zone.
This can be verified by noting 
 that the present nodal line is almost  an ellipse with 
 major axis $a \simeq 0.44 $ and minor axis $b \simeq 0.086$. 
The ellipse length $\tilde{L}$ is given by 
 $\tilde{L} = 4aE(k)$ with $k^2= 1 -(b/a)^2$, 
 $E(k)$ being the complete elliptic integral of the second kind defined by 
$ E(k) = \int^{\pi/2}_0 [ 1- k^2 \sin^2 x]^{1/2} d x$. 
 Since  $E(k) \simeq 1.05$ with $k^2 = 0.96$ for the present case,
 we obtain $\tilde{L} = 1.85 \simeq L$, which well reproduces the above numerical 
result. 
Furthermore, the area of the nodal line (ellipse) is estimated as  
 $\pi a b = \pi \tilde{L}^2 (1-k^2)^{1/2}/(16 E(k)^2) \simeq 0.118$, which 
 is  much smaller than 1, corresponding to the area of the first Brillouin zone.

\begin{figure}
  \centering
\includegraphics[width=6cm]{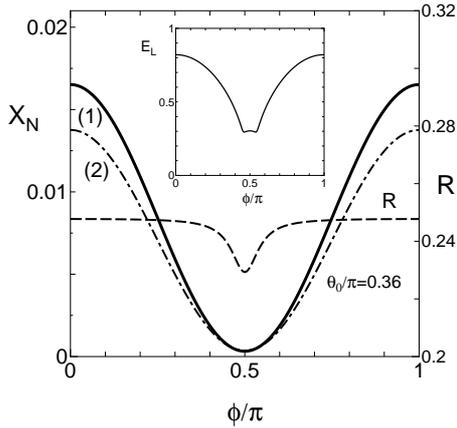}
    \caption{
(Color online)
Orbital magnetic susceptibility $\chi_N$ 
under  magnetic  field $\bm{B}$ $(= B \ve_{B})$  
  applied in the plane of the nodal line 
   and  given by Eq.~(\ref{eq:x26}).
The phase  $\phi$ is the rotation angle around a normal 
vector $\ve_{np} [ =(\sin \theta_0, 0, \cos \theta_0)]$  perpendicular to the plane of the nodal line, 
where  $\ve_{B} =  (-\cos\theta_0, 0, \sin \theta_0)$ for $\phi=0$.   
 Line (1) corresponds to a half-filled band 
 and  line (2) denotes a  case of hole  doping, 
 where the chemical potential is given 
   by 
  $\delta \mu$ = 0 (1) and  -0.003 (2). 
   $\delta \mu = \mu - \mu_0$ and  $\mu_0$ =  -0.0041.
 The dashed line denotes  $R$  defined by 
 $R = \chi_N / <(\ve_{B} \cdot \ve_{1})^2>$, 
which is almost constant except in the region 
 close to  the minimum (at $\phi/\pi = 0.5$).  
In the inset, $E_L$ is shown as a function of $\phi/\pi$ 
for a magnetic field  perpendicular to the normal vector $\ve_{np}$. 
 }
\label{fig6}
\end{figure}

Next,  to examine the plane of the nodal line, we calculate a unit 
vector $\ve_{np}$  
 perpendicular to the plane. 
Since the present nodal line is  non-coplanar,
 the condition $\ve_{np}\cdot \ve_1 = 0$ is not always satisfied 
 on the line. 
Therefore, we determine $\ve_{np}$ to give the  minimum 
   $<|\ve_{np}\cdot \ve_{1}|>$,
 where  $\ve_1$ is given 
  by Fig.~\ref{fig3}(a). 
 Because of  the mirror symmetry at $k_y=0$,
 we calculate $\ve_{np}$  in the form 
of $\ve_{np} = (\sin \theta , 0, \cos \theta)$, 
where $\theta$ denotes the angle  between 
$\ve_z$ and $\ve_{np}$. 
The minimum  $<|\ve_{np}\cdot \ve_{1}|>$ 
  is obtained at $\theta = \theta_0 \simeq 0.36 \pi$,  
   where $<|\ve_{np}\cdot \ve_{1}|> \simeq 0.029$. 
 This means that the deviation of the nodal line from te plane is  
 moderately small.

\subsection{Orbital magnetic susceptibility}

   As one of the characteristic physical quantities, we calculate the orbital susceptibility for   the nodal line semimetal. 
  It is well known that the two-dimensional massless Dirac electrons (or electrons in graphene) give a delta-function-like
  orbital magnetic susceptibility,~\cite{McClure,Fukuyama1971,Fukuyama2007} 
\begin{equation}
\chi_{\rm GR} = -\frac{e^2 v^2}{3\pi^2 \Gamma_0}\frac{1}{1+X_0^2},
\end{equation}
where $v$ is the velocity of the Dirac cone, $X_0=\mu/\Gamma_0$, and the relaxation rate $\Gamma_0$ 
has been introduced phenomenologically. Note that in the limit of $\Gamma_0\rightarrow 0$, 
$1/\Gamma_0 (1+X_0^2)$ becomes $\pi\delta(\mu)$. 
The effect of tilting on the magnetic susceptibility was studied previously.\cite{Kobayashi2008,Ogata_note} Here we use
\begin{equation}
\chi_{\rm tilting} = - \frac{e^2 v^2}{3\pi^2 \Gamma_0}\frac{1}{1+X_0^2}(1-\eta^2)^{3/2}.
 \label{TiltingChi}
\end{equation}

In this study, we estimate the orbital magnetic susceptibility $\chi$ for the present nodal line, assuming that 
$\chi$ is given by the sum of $\chi_{\rm tilting}$ over the nodal line. 
To study the angle dependence of $\chi$, 
the direction of the magnetic field is chosen to be within the nodal plane. 
For this purpose, 
 the unit vector of  the magnetic field, $\ve_{B}=\bm{B}/B $, is taken  as
\begin{eqnarray}
 \ve_{B} = (- \cos \phi \cos \theta_0, \sin \phi, \cos \phi \sin \theta_0 ) \; ,  \label{eq:x26}
 \end{eqnarray}
so that $\bm B$ is always perpendicular to 
$\ve_{np} = (\sin \theta_0, 0, \cos \theta_0)$.
The phase $\phi$ represents the angle of $\bm B$  
 and is chosen such that $\ve_{B} =  (-\cos\theta_0, 0, \sin \theta_0)$ when $\phi=0$. 
Therefore, the phase $\phi=0$ corresponds to the case where $\bm B$ is parallel to the 
tangential direction at  Dirac point  (I). 
In the same way, $\phi=\pi/2$ corresponds to the case where $\bm B$ is tangential 
at  Dirac point (II).

Since the principal axes for the Dirac cone are located in the $\ve_2$-$\ve_3$ plane, 
the effective magnetic field is considered to be $(\ve_B \cdot \ve_1)B$. 
Taking account of  the variation of velocities   
  and $\delta E_{\rm D}$ along the nodal line, 
$\chi$ is written as
\begin{subequations}
\begin{eqnarray}
 & &\chi =  -\frac{e^2 <V_+V_->L}{3\pi^2 \Gamma_0} \chi_N \; ,
 \label{eq:x27a}
 \\
 & & \chi_N = 
 \left<   (\ve_B \cdot \bm{e}_1)^2 \frac{V_+V_-}{<V_+V_->} 
\frac{1}{1+\tilde{X}^2} 
 (1 - \eta^2)^{3/2} 
   \right>   \; , 
 \nonumber \\
\label{eq:x27b}
\end{eqnarray}
where $\tilde{X} = (\mu - \delta E_D)/\Gamma_0$   with $\Gamma_0 = 0.001$.
Quantities  $V_\pm, \eta$, and  $\tilde{X}$
vary along the 3D nodal line [Fig.~\ref{fig2}(a)].
Figure \ref{fig6} shows $\chi_N$ [line (1)] as a function of $\phi$, where 
$\chi_N$ is maximum,  $\chi_N \simeq$  0.0165, 
 at $\phi/\pi$ = 0 and 1, and   minimum,    $\chi_N \simeq$ 0.0003,  
 at $\phi/\pi$ = 0.5. 
This means that 
the maximum (minimum)  $\chi$ occurs when $\bm B$ is parallel to $\ve_{1}$
   at Dirac point (I) 
(Dirac point (II)). 
 The reason why $\chi_N$ is minimum at $\phi/\pi$ = 0.5 is that the Dirac cone near 
  Dirac point (II) is fairly distorted, as seen from  Figs.~\ref{fig3}(b),
\ref{fig3}(c), and \ref{fig4}. 
The experimental observation of 
 such an extremum   will be useful in finding 
the direction of the principal axis of the nodal plane.

We examine $\chi_N$ using a quantity $R$ given by  
\begin{eqnarray}
 R = \frac{\chi_{N}}{<(\ve_{B} \cdot \ve_{1})^2>} \; ,
 \label{eq:x27c}
 \end{eqnarray}
\end{subequations}
which is shown as a function of $\phi$ in Fig.~\ref{fig6}. 
The quantity $R$ gives an estimation of the average of orbital magnetic 
 susceptibility 
 along the  nodal line without considering  
  the  weight,  $ (\ve_{B} \cdot \ve_{1})^2$.
We obtain  $R \simeq 0.2477$ at  $\phi/\pi$ = 0 and 
 $R \simeq 0.2293$ at $\phi/\pi$ =0.5. 
Since the variation of $R$ is  small compared with  
 that of $\chi_N$,    the $\phi$ dependence of $\chi_N$ 
 is essentially determined by $(\ve_{B} \cdot \ve_{1})^2$, i.e., 
 the geometric property of the nodal line. 
The  average of the respective quantities in Eq.~(\ref{eq:x27b})
 is  estimated as 
$<V_+V_->$ = 0.2462, 
$\sqrt{<\eta^2>}$ = 0.528, 
$<1/(1+\tilde{X}^2)>$ = 0.364,  and  
$<(\ve_{B}\cdot \ve_1)^2>$ = 0.0665 (for $\phi/\pi$ = 0),  
 and $<( 1 - \eta^2)^{3/2}>$ = 0.614.
 Furthermore, we note that $( 1 - <\eta^2>)^{3/2}\simeq 0.612$. 
 Using the average quantity, we obtain 
 $<1/(1+\tilde{X}^2)> <( 1 - \eta^2)^{3/2}> \simeq$ 
    0.223, which is slightly  smaller than $R$ in Fig.~\ref{fig6}.
Such an enhancement of $R$ compared with the product of the average quantities  comes from a combined effect of  
 $(\ve_{B}\cdot \ve_1)^2$, $V_+V_-$,  $1/(1+\tilde{X}^2)$, and 
 $( 1 - \eta^2)^{3/2}$. 
 
Here, we briefly mention 
 the effect of  carrier doping, which is given by 
the variation of chemical potential, i.e., 
$\delta \mu (= \mu - \mu_0)$ with $\mu_0$= - 0.0041.
In Fig.~\ref{fig6},  $\chi_N$ for  hole  doping is  shown 
 by  line (2) with $\delta \mu$ = -0.003, 
 which is smaller than that of  line (1). 
Note that,the  $\phi$-dependence of $\chi_N$ does not change qualitatively. 
 We find that,  with decreasing $\delta \mu$,
 $\chi_N$ first becomes maximum at $\delta \mu$ = -0.0018 and decreases rapidly, whereas 
 $\chi_N$ decreases monotonically for increasing $\delta \mu ( >0)$. 
 Such  asymmetry comes from that of the DOS (Fig.~\ref{fig5}) and 
  the variation in the energy $\delta E_{\rm D}$ on the nodal line in Fig.~\ref{fig2}(a).  

Finally, we discuss the average  Landau level of  the nodal line 
 for  a magnetic field $\bm{B} = B \ve_{B}$  
 in  the nodal plane. 
   The  $N$th  Landau level is given by 
  $ E_{\pm N} = \pm \sqrt{2Ne \hbar B ( 1 - \eta^2)^{3/2}v^2}$
\cite{Morinari2009}. When we take the average over the nodal line, we obtain 
  $ E_{\pm N} = \pm \sqrt{2Ne \hbar B}<\sqrt{V_+V_-}> E_L + E_0$,
  where 
\begin{eqnarray}
 E_{L} = \frac{  < \sqrt{V_+V_-} (1 - \eta^2)^{3/4}
 \sqrt{|\ve_{B} \cdot \ve_{1}|} > } { <\sqrt{V_+V_-} >}\; .
\label{eq:x28}
 \end{eqnarray}
 The quantity $E_0$ is the average 
 energy of the zeroth Landau level measured 
 from the chemical potential, which is given by 
 $<\delta E_{D}(\bm{k}_0) - \mu> \simeq -0.0002$. 
 The $\eta$-dependence of Eq.~(\ref{eq:x28})
   has a common feature with that of  Eq.~(\ref{TiltingChi}), 
 suggesting  that 
 the effect of the  magnetic field on the Landau orbit 
 is reduced by a factor of   $(1 - \eta^2)^{3/2}$ \cite{Baskaran}.
 In the inset of Fig.~\ref{fig6}, $E_{L}$ 
     is shown as a function of $\phi$  
      for a magnetic field  perpendicular to 
 $ \ve_{np} =(\sin \theta_0, 0, \cos \theta_0)$.
Note that $E_L$ in  the plane perpendicular to 
 $(\sin \theta, 0, \cos \theta)$ 
 is maximum at $\theta = \theta_0$.
 The behavior in  the inset  is similar to that of  $\chi_N$ but 
 differs at  around  $\phi/\pi = 1/2$.
$|E_{\pm 1}|$  will be identified 
 when  $|E_{\pm 1}|$ is  larger than the variation 
  of $\delta E_{D}$. 
 For a two-dimensional organic conductor,\cite{Tajima2010_JPSJ} 
  it has been claimed that  the peak of 
 the temperature dependence of interlayer   
 longitudinal magnetoresistance  
  is associated with 
 the energy  separation between the $N = 0$ and $N = \pm 1$ Landau levels.
 Therefore, if the $\phi$-dependence of  $|E_{\pm 1}|$ is estimated from 
  such a magnetoresistance experiment, we can find the direction of the 
 principal axis of the nodal line.  
 
\section{Summary}
We examined an effective Hamiltonian of a two-band model, 
    which describes the  Dirac cone  close to the nodal line of 
 a molecular conductor [Pd(dddt)$_2$] with a half-filled band. 
  The energy with  a dispersion perpendicular to the  nodal line  
 was evaluated using the Dirac points obtained by the DFT calculation. 
   The energy difference between the conduction and valence bands 
 was calculated to obtain the principal axes and corresponding 
velocities, $V_+$ and $V_-$, which   
  rotate   along the nodal line. 
  Furthermore, 
 the effect of  tilting  on the Dirac cone was examined, where 
  the mutual relationship  between  the principal axis and the tilting direction was clarified. 
   The Dirac cone obtained 
   by  varying  the nodal line 
     gave reasonable energies,   
 since the density of states showing 
 the   characteristics of the nodal line semimetal 
    was i compatible with that obtained by  the DFT calculation.
  The  determination of the tilting  axis of the respective Dirac cone 
    in terms of the original momentum space $\bm{k}$ 
      is useful for calculaing an response to 
       the external field with arbitrary direction.
As  an example, we demonstrated the angular dependence of 
orbital magnetic susceptibility $\chi_N$,  where  the magnetic field was
  applied in the plane of the nodal line.
Finally, we noted  that the present method of deriving the effective Hamiltonian 
 for the Dirac cone  could be applied to other systems of a nodal line 
 with an inversion  symmetry.  

\acknowledgements
One of the authors (Y.S.) thanks   A. Yamakage for useful discussions. 
This research was funded by JSPS Grants-in-Aid for Scientific Research 
No.~16H06346, 16K17756, 18H01162, 18K03482, 19K03720, and  19K21860,
   and JST CREST Number JPMJCR18I2, Japan.
Computational work was performed under the Inter-university Cooperative Research Program and the Supercomputing Consortium for Computational Materials Science of the Center for Computational Materials Science of the Institute for Materials Research (IMR), Tohoku University (Proposasl No. K18K0090 and 19K0043). The computations were mainly carried
out using the computer facilities of ITO at Kyushu University, MASAMUNE-IMR at Tohoku University, and HOKUSAI-GreatWave at RIKEN.
%


\end{document}